\documentclass[conference]{IEEEtran}
\IEEEoverridecommandlockouts

\usepackage{cite}
\usepackage{amsmath,amssymb,amsfonts}
\usepackage{algorithmic}
\usepackage{graphicx}
\usepackage{textcomp}
\usepackage{xcolor}
\usepackage{booktabs}
\usepackage{makecell}
\usepackage{subcaption}
\def\BibTeX{{\rm B\kern-.05em{\sc i\kern-.025em b}\kern-.08em
    T\kern-.1667em\lower.7ex\hbox{E}\kern-.125emX}}
\begin{document}

\title{Q-DIVER: Integrated Quantum Transfer Learning and Differentiable Quantum Architecture Search with EEG Data}

\IEEEaftertitletext{\centering \vspace*{-2.0\baselineskip} \footnotesize \textsuperscript{*} These authors contributed equally \vspace{1.5\baselineskip}}

\author{\IEEEauthorblockN{Junghoon Justin Park*}
\IEEEauthorblockA{
\textit{Seoul National University}\\
Seoul, Korea \\
utopie9090@snu.ac.kr}
\and
\IEEEauthorblockN{Yeonghyeon Park*}
\IEEEauthorblockA{
\textit{Seoul National University}\\
Seoul, Korea \\
mandy1002@snu.ac.kr}
\and
\IEEEauthorblockN{Jiook Cha}
\IEEEauthorblockA{
\textit{Seoul National University}\\
Seoul, Korea \\
connectome@snu.ac.kr}
}

\maketitle

\begin{abstract}
Integrating quantum circuits into deep learning pipelines remains challenging due to heuristic design limitations. We propose Q-DIVER, a hybrid framework combining a large-scale pretrained EEG encoder (DIVER-1) with a differentiable quantum classifier. Unlike fixed-ansatz approaches, we employ Differentiable Quantum Architecture Search to autonomously discover task-optimal circuit topologies during end-to-end fine-tuning. On the PhysioNet Motor Imagery dataset, our quantum classifier achieves predictive performance comparable to classical multi-layer perceptrons (Test F1: 63.49\%) while using approximately \textbf{50$\times$ fewer task-specific head parameters} (2.10M vs. 105.02M). These results validate quantum transfer learning as a parameter-efficient strategy for high-dimensional biological signal processing.

\end{abstract}


\section{Introduction}
Recent progress in quantum machine learning has demonstrated that quantum circuits can be integrated into classical deep learning pipelines in various ways. These integration-focused approaches have been instrumental in establishing the feasibility of classical--quantum hybrid models. However, they leave a more fundamental question open: at this stage, which functional components of a classical learning pipeline should quantum models meaningfully replace rather than merely augment?

From a practical standpoint, classical machine learning pipelines are already highly mature, while near-term quantum devices operate under significant constraints. As a result, replacing entire classical systems with quantum models---or introducing quantum components without a clearly defined functional role---is neither necessary nor well-motivated.
These considerations call for a more deliberate and structurally informed approach to integrating quantum models, in which their functional role within the classical pipeline is explicitly specified rather than implicitly assumed. In particular, this motivates a systematic exploration of where and how quantum components should be incorporated, rather than relying on ad hoc circuit choices.

This perspective aligns with a long-standing view in classical machine learning that learning systems can be decomposed into representation learning and task-specific decision making~\cite{Yosinski2014,Bengio2013}. More recently, the emergence of foundation models has elevated representation reuse into a central design paradigm, explicitly treating pretrained representations as reusable assets across downstream tasks~\cite{Bommasani2021}. On a technical level, this paradigm is fundamentally enabled by transfer learning, where large-scale pretraining decouples representation learning from task-specific decision making.

Among the many application domains that have adopted this paradigm, electrophysiological signal analysis has applied large-scale self-supervised pretraining to learn transferable spatiotemporal representations~\cite{Wang2024EEGPT,Jiang}.
However, EEG signals exhibit substantial inter- and intra-subject variability and are non-stationary, leading to pronounced distribution shifts across subjects and sessions that challenge downstream generalization~\cite{Apicella2024,Wang2024Cbramod}.

Recent studies have suggested that downstream readout mechanisms can play an important role in EEG transfer learning, and that lightweight or generic classifiers may not fully exploit pretrained representations~\cite{Liang2024,Jeon2025}.
 Nevertheless, the readout stage is typically treated as part of a broader optimization pipeline rather than examined as a primary object of analysis.
In this work, we investigate the downstream readout stage as a standalone modeling component and explore the use of quantum models as alternative readout mechanisms acting on fixed classical representations. Motivated by prior studies on quantum time-series transformers, we repurpose this architecture as a quantum readout module. To avoid attributing potential performance gains to arbitrary circuit choices, we adopt differentiable quantum architecture search (DiffQAS) as a systematic optimization framework, enabling controlled analysis of the structural role of the quantum readout.

\section{Background}

\subsection{Quantum Transfer Learning} 
Transfer learning considers settings in which the source and target domains and/or tasks differ \cite{Pan2009}. The transfer learning paradigm has been extended to hybrid
classical--quantum architectures~\cite{Mari2020}, where transfer is defined by the reuse
of representations across stages, independent of whether the source or target models are
classical or quantum. Among various settings, the classical--quantum configuration—where a classical backbone provides representations to a quantum readout—has received particular attention in the Near-Term Intermediate Scale Quantum (NISQ) era due to its practicality~\cite{Mari2020}.

Recent quantum transfer learning studies explore different ways of realizing
transfer within this hybrid framework.
Tseng \emph{et al.}~\cite{Tseng2025} formulate transfer learning directly at the level of the variational quantum circuit (VQC), modeling domain adaptation as a one-step algebraic estimation of parameter shifts under a fixed circuit and measurement.
In contrast, many hybrid approaches freeze a deep classical network pretrained on
large-scale datasets (e.g., ImageNet) as a feature extractor and train a VQC as a
task-specific classification head on the target domain~\cite{Mari2020, Khatun2025}.


\subsection{DIVER-1}

Recent advances in electrophysiology foundation models show that large-scale pretraining on heterogeneous neural recordings yields representations that generalize across subjects, recording setups, and downstream tasks. DIVER-1 \cite{Han2025} exemplifies this through scale, training strategy, and architectural design.

DIVER-1 is pretrained on a diverse corpus of EEG and iEEG data from over 17,700 subjects, spanning substantial inter-subject and cross-modality variability \cite{Han2025}. Its data-constrained scaling analysis shows that representation quality depends not only on parameter count, but also on training duration and data diversity, with smaller models trained longer sometimes outperforming larger models trained briefly.

Prior work suggests that transfer gains in data-limited scientific domains may reflect overparameterization or training dynamics rather than genuine feature reuse \cite{Raghu2019}. The scaling behavior observed in DIVER-1 supports the view that transferable electrophysiological representations cannot be attributed to model size alone.

Architecturally, DIVER-1 incorporates permutation equivariance and support for heterogeneous sensor configurations to promote cross-subject and cross-setup robustness \cite{Han2025}. These choices reduce sensitivity to electrode ordering and recording systems, making DIVER-1 a suitable backbone for studying how pretrained electrophysiological representations are leveraged by downstream prediction heads.



\subsection{Quantum Time-series Transformer}
To effectively capture long-range temporal dependencies within the high-dimensional EEG embeddings while minimizing parameter overhead, we employ the \textit{Quantum Time-series Transformer} (QTSTransformer) \cite{Park2025}. Unlike classical transformers that rely on computationally intensive quadratic attention mechanisms ($\mathcal{O}(L^2)$), the QTSTransformer leverages quantum mechanical properties to process sequence data with polylogarithmic complexity ($\mathcal{O}(\text{polylog}(L))$). The model's architecture is defined by four distinct operational stages:

\subsubsection{Unitary Temporal Embedding}
The first stage maps classical temporal data into the quantum Hilbert space. Let $\mathbf{X} = \{x_0, \dots, x_{n-1}\}$ denote the sequence of latent feature vectors extracted by the classical backbone. Each feature vector $x_j$ at time-step $j$ is mapped to a unique Variational Quantum Circuit (VQC), which defines a specific unitary transformation $U_j(\theta_j)$. This process creates a quantum representation where the temporal information is encoded directly into the operation of the quantum gate sequence.

\subsubsection{Time Sequence Mixing via LCU}
To model temporal correlations, we employ the \textit{Linear Combination of Unitaries} (LCU) primitive. This serves as a quantum-native attention mechanism. Instead of explicitly computing a pairwise attention matrix, the model prepares a control register in a superposition state defined by coefficients $\alpha$. This control state directs the simultaneous application of the sequence unitaries $\{U_0, \dots, U_{n-1}\}$ onto a target register. The resulting operation creates a weighted superposition operator $M$, defined as:

\begin{equation}
    M = \sum^{n-1}_{j=0} e^{i\gamma_j} |a_j|^2 U_j
\end{equation}

\noindent where $e^{i\gamma_j}$ represents the phase and $|a_j|^2$ represents the magnitude of the attention weight for the $j$-th time step. This allows the model to "attend" to multiple time-steps simultaneously through quantum interference.

\subsubsection{Non-linearity via QSVT}
To introduce the non-linearity required for complex decision boundaries, we utilize the \textit{Quantum Singular Value Transformation} (QSVT). While classical transformers rely on activation functions like Softmax or GELU, the QTSTransformer applies a polynomial transformation $P_c$ to the singular values of the mixed operator $M$. The transformed operator takes the form:

\begin{equation}
    P_c(M) = c_d M^d + c_{d-1} M^{d-1} + \cdots + c_1 M + c_0 I
\end{equation}

\noindent where $\{c_0, \dots, c_d\}$ are trainable polynomial coefficients and $d$ is the degree of the polynomial. This transformation allows the model to capture richer, higher-order interactions between different time-steps without collapsing the quantum state.

\subsubsection{Readout and Classical Processing}
In the final stage, the transformed quantum state is measured to extract classical features via Pauli expectation values. These values are subsequently processed by a compact classical feed-forward neural network to produce the final downstream prediction (e.g., classification of motor imagery tasks).

This architecture serves as the decision head of our hybrid framework. By processing the pre-trained features through this quantum-native attention mechanism, we achieve high representational power with significantly fewer trainable parameters than a comparable classical multi-layer perceptron (MLP).

\subsection{Differentiable Quantum Architecture Search}

Performance of variational quantum circuits can be highly sensitive to architectural design and initialization, making it difficult to disentangle intrinsic quantum limitations from incidental circuit choices \cite{McClean2018, Holmes2022}. To enable principled evaluation and task-specific adaptation, we adopt a differentiable quantum architecture search (DiffQAS) framework \cite{DiFFQAS}, which relaxes discrete circuit design into a continuous, end-to-end optimizable formulation.

\subsubsection{Search Space Construction}

We construct a modular search space in which a quantum circuit $C$ consists of $L$ sequential units $\mathcal{S}_1, \dots, \mathcal{S}_L$. Each unit $\mathcal{S}_l$ is selected from a predefined candidate library $\mathcal{B}_l$. The total number of possible circuit configurations is

\begin{equation}
    N = \prod_{l=1}^{L} |\mathcal{B}_l|.
\end{equation}

The candidate library includes different entanglement patterns and parameterized single-qubit rotation gates (e.g., $R_x, R_y, R_z$), allowing exploration of varying expressivity and correlation structures.

\subsubsection{Differentiable Optimization Framework}

Rather than selecting a single discrete architecture, we assign a learnable structural weight $w_j$ to each candidate configuration $\mathcal{C}_j$. Let $f_{\mathcal{C}_j}(\mathbf{x}; \theta_j)$ denote the output of candidate $\mathcal{C}_j$ with parameters $\theta_j$. The effective model output is defined as the weighted ensemble

\begin{equation}
    f_{\text{ens}}(\mathbf{x}) = \sum_{j=1}^{N} w_j f_{\mathcal{C}_j}(\mathbf{x}; \theta_j).
\end{equation}

This continuous relaxation enables joint optimization of circuit parameters and architecture.

\subsubsection{Training and Discretization}

We jointly optimize the variational parameters $\Theta = \{\theta_1, \dots, \theta_N\}$ and structural weights $\mathcal{W} = \{w_1, \dots, w_N\}$ by minimizing

\begin{equation}
    \min_{\Theta, \mathcal{W}} \mathcal{L}(f_{\text{ens}}(\mathbf{x}), y).
\end{equation}

After convergence, the final discrete architecture is obtained by selecting the candidate with the largest structural weight:

\begin{equation}
    \mathcal{C}_{\text{final}} = \arg\max_{j} w_j.
\end{equation}

In our implementation, DiffQAS is applied in a factorized manner by optimizing the Timestep Modeling and QFF blocks separately and composing the selected candidates.

\section{Q-DIVER}
Our Q-DIVER is a hybrid framework that integrates the classical and quantum components into a unified pipeline designed for data-efficient fine-tuning on downstream EEG tasks.

\subsection{Q-DIVER Hybrid Architecture}
\begin{figure}
    \centering
    \includegraphics[width=0.75\linewidth]{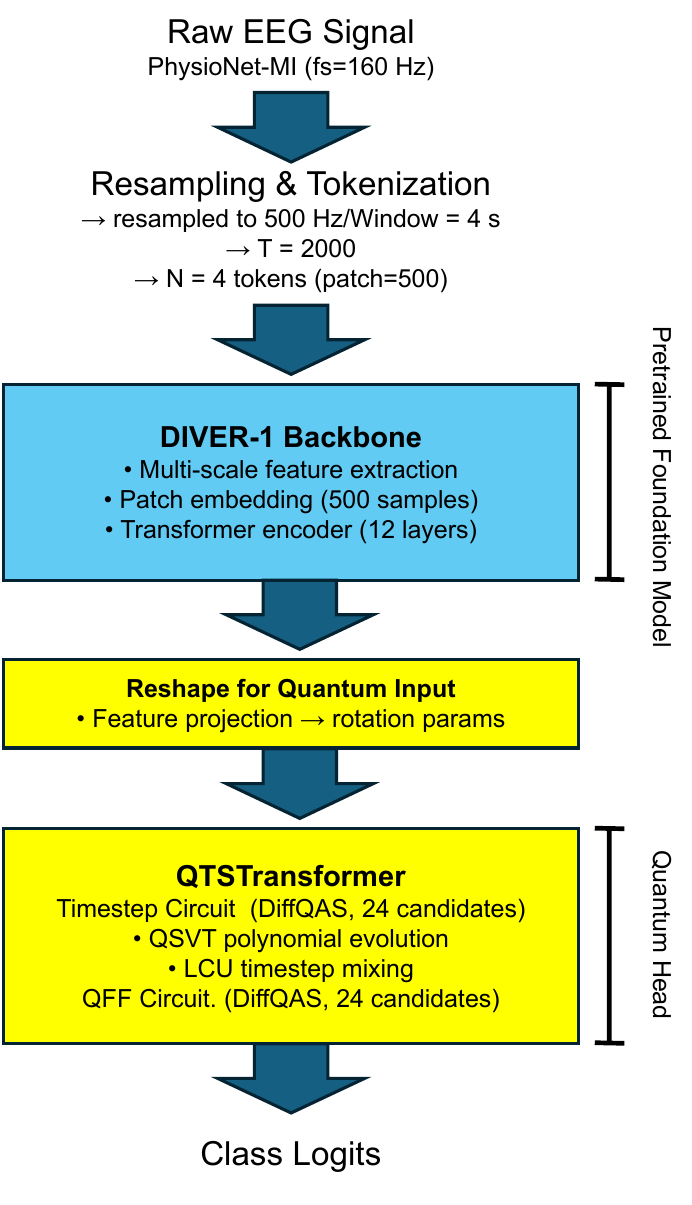}
    \caption{Overview of the Q-DIVER hybrid architecture. Raw EEG signals are processed by the pretrained DIVER-1 backbone and passed to a QTSTransformer-based quantum head. DiffQAS is applied to the Timestep and QFF circuit blocks (24 candidates each).} 
    \label{fig_Q-DIVER}
\end{figure}

The Q-DIVER pipeline effectively bridges the gap between high-dimensional classical data and the quantum feature space. As illustrated in Fig. \ref{fig_Q-DIVER}, the architecture operates as a two-stage transfer learning system.

\subsubsection{The Hybrid Pipeline}
The data flow proceeds sequentially through the classical backbone and the quantum classification head:

\begin{enumerate}
    \item \textbf{Input Processing:} The pipeline accepts raw EEG signals denoted as $X \in \mathbb{R}^{B \times C \times T}$, where $B$ is the batch size, $C$ is the number of channels, and $T$ represents time points.
    
    \item \textbf{Classical Backbone (DIVER-1):} The pre-trained DIVER-1 encoder processes the input $X$ to extract high-level spatio-temporal representations. Unlike standard transfer learning where the backbone is often frozen, we perform end-to-end fine-tuning, allowing the backbone to adapt its latent space specifically for the quantum manifold. The output is a latent representation $H_{\text{class}} \in \mathbb{R}^{B \times N \times d}$, where $N$ is the sequence length (patches) and $d$ is the feature dimension.
    
    \item \textbf{Quantum Projection Interface:} To encode the classical features into the quantum circuit, we employ a learnable projection layer with regularization. For each timestep $t$, the classical feature vector $x_t \in \mathbb{R}^{d_{\text{feat}}}$ is mapped to quantum rotation parameters $\theta_t \in [0, 1]^{n_{\text{params}}}$ via:
    \begin{equation}
        \theta_t = \sigma(W \cdot \text{Dropout}(x_t) + b)
    \end{equation}
    where $\sigma(\cdot)$ is the sigmoid function, ensuring parameters remain within the normalized range. These parameters are then applied to the quantum gates as rotation angles scaled by $\pi/2$:
    \begin{equation}
        R_\alpha(\theta_{t,i}) = \exp\left(-i \cdot \frac{\pi \cdot \theta_{t,i}}{2} \cdot \sigma_\alpha\right)
    \end{equation}
    where $\alpha \in \{X, Y, Z\}$ corresponds to the selected Pauli rotation gate.
    
    \item \textbf{Quantum Classification Head:} The projected features drive the QTSTransformer. The quantum head applies the optimal circuit ansatz discovered via DiffQAS to perform unitary evolution using LCU and QSVT. Finally, quantum measurement (expectation values of Pauli operators) yields the logits $Y_{\text{out}} \in \mathbb{R}^{B \times K}$, where $K$ is the number of classes.
\end{enumerate}


\subsection{Parameter Breakdown and Efficiency}

The QTSTransformer head introduces substantially fewer task-specific parameters 
than a classical dense MLP head applied to flattened spatio-temporal features. 
The DIVER-1 backbone contains 51.36M parameters. In the end-to-end fine-tuning 
setting, both the backbone and the classifier head are trainable.

\begin{table}[ht]
\centering
\caption{Parameter breakdown comparison between DIVER+MLP and DIVER+QTSTransformer (PhysioNet-MI, 4-second trial setting, 8 qubits). Both models are trained end-to-end.}
\label{tab:param_compare_breakdown}
\vspace{0.2cm}
\setlength{\tabcolsep}{3.5pt} 
\begin{tabular}{@{}lcc@{}}
\toprule
\textbf{Component} 
& \textbf{DIVER+MLP} 
& \textbf{DIVER+QTSTransformer} \\
\midrule
DIVER-1 Backbone & 51.36M & 51.36M \\
Feature Projection & --- & 2.10M \\
Classifier & 105.02M & 143 \\
\midrule
\textbf{Total} & \textbf{156.38M} & \textbf{53.46M} \\
\bottomrule
\end{tabular}
\end{table}

As shown in Table~\ref{tab:param_compare_breakdown}, the DIVER+MLP configuration 
contains 156.38M trainable parameters, whereas DIVER+QTSTransformer contains 
53.46M parameters under end-to-end training. This corresponds to an overall 
parameter reduction of approximately $\mathbf{2.9\times}$ at the full-model level.

The difference becomes substantially more pronounced when focusing only on the task-specific classifier head. The classical MLP head introduces 105.02M parameters, while the QTSTransformer head requires only 2.10M parameters, yielding an approximate $\mathbf{50\times}$ reduction in head-specific trainable parameters.

Importantly, in the quantum configuration, the majority of head parameters reside in the classical feature projection layer(2.10M), while the quantum circuit itself introduces only 43 trainable parameters. This corresponds to approximately 0.002\% of the head parameters and about 0.00008\% of the total model parameters, confirming that the quantum component contributes a negligible fraction of the overall parameter count.

\section{Experiment}
\subsection{Pretraining}
The backbone encoder used in this study follows the DIVER-1 self-supervised pretraining procedure~\cite{Han2025} and is pretrained on a large-scale corpus of EEG and iEEG recordings aggregated from multiple publicly available datasets.

Raw EEG and iEEG signals were standardized following the DIVER-1 preprocessing protocol: resampling to 500~Hz and minimal filtering with a 0.3--0.5~Hz high-pass and a 60~Hz notch filter, without low-pass filtering. Each training sample consisted of a 30-second continuous window tokenized into temporal patches of either 0.1~s or 1.0~s duration.

The pretraining data include intracranial EEG datasets such as AJILE12~\cite{Peterson2022} and the Penn Electrophysiology of Encoding and Retrieval Study (PEERS)~\cite{Michael2023}, a self-collected iEEG dataset from epilepsy patients, as well as large-scale scalp EEG datasets including the Temple University Hospital EEG Corpus (TUEG)~\cite{Obeid2016}, the Healthy Brain Network EEG dataset (HBN-EEG)~\cite{Shirazi2024}, and the Nationwide Children’s Hospital Sleep DataBank (NCHSDB)~\cite{Lee2022}. These datasets span diverse recording modalities and subject populations, supporting robust representation learning across heterogeneous electrophysiological signals.

The pretraining objective operates on multiple signal representations. In addition to reconstructing the raw time-domain signal, masked patches are reconstructed in complementary domains such as spectral representations, and the corresponding errors are aggregated into a single loss. To improve robustness to heterogeneous recording configurations, random subsampling was applied during pretraining: for each 30-second window, up to 32 channels and up to 30 temporal patches were selected. For the 0.1~s patch configuration, the number of patches was capped at 30.

Unmasked patches are processed by a lightweight patch-wise convolutional neural network to extract local temporal features and project them into token representations compatible with the transformer backbone. The resulting pretrained encoder is reused for downstream tasks with different prediction heads.

\subsection{Downstream Task}
We evaluate Q-DIVER on a publicly available EEG dataset not used during pre-training: 
the \textit{PhysioNet Motor Imagery (PhysioNet-MI)} dataset~\cite{Goldberger2000}. PhysioNet-MI includes EEG recordings from 109 subjects acquired with 64 channels at 160~Hz. The dataset comprises four motor imagery classes (left fist, right fist, both fists, both feet), and we evaluate a 4-class motor imagery classification task.







\subsection{QAQC and Preprocessing}

Quality assessment and minimal preprocessing followed the same protocol used in DIVER-1 pretraining, including amplitude normalization and clipping-based rejection.

\subsection{Experimental Setting}
We evaluate the framework on the downstream EEG dataset (PhysioNet-MI) using a rigorous fine-tuning protocol. 
Raw EEG recordings (originally sampled at 160 Hz for PhysioNet-MI) were re-referenced and minimally filtered using a 0.3–0.5 Hz high-pass filter and a 60 Hz notch filter. 
All signals were then resampled to 500 Hz to maintain consistency with the DIVER-1 pretraining configuration. 
Each trial in the PhysioNet-MI dataset corresponds to a 4-second motor imagery interval. 
We directly used these trial segments as input samples without introducing additional temporal windowing.

\subsubsection{Hyperparameters and Configuration}
The hybrid model is trained end-to-end using the Cross-Entropy loss function. We utilize the AdamW optimizer to update both the classical backbone weights and the quantum circuit parameters. The specific hyperparameters used for fine-tuning are: batch size 32, learning rate $5e^{-5}$, weight decay $1e^{-2}$, and dropout 0.1 (applied to the classical projection layer).

\subsubsection{DiffQAS Search Space}

Instead of relying on stochastic Monte Carlo sampling~\cite{DiFFQAS}, we employ a structured factorized search over the same candidate space. Each of the two circuit blocks (Timestep Modeling and QFF) is optimized independently as a 24-way softmax mixture, introducing $24 + 24 = 48$ learnable structural weights while implicitly representing all $24 \times 24 = 576$ joint configurations. 

Given the moderate search size, all candidates within each block are evaluated deterministically at every step. After a short warmup with uniform mixing ($1/24$), structural weights are optimized via softmax relaxation, and the final architecture is obtained by argmax selection in each block.

\begin{table}[]
    \centering
    \caption{Search Space for Differentiable Quantum Architecture Search}
    \begin{tabular}{lc}
    \toprule
    \textbf{Component} & \textbf{Options} \\
    \midrule
       Initialization  & $\{ \text{Hadamard}, \text{None} \}$ \\
       Entanglement & \makecell{$\{ \text{Linear CNOT}, \text{Ring CNOT}$, \\ $\text{CRX Forward}, \text{CRX Backward} \}$} \\
       Variational Gates & $\{ \text{RX}, \text{RY}, \text{RZ} \}$ \\
       \bottomrule
    \end{tabular}
    \label{table_DiffQAS_space}
\end{table}


\section{Results}
We present the PhysioNet-MI experiment results demonstrating the model's classification performance, architectural convergence, and parameter efficiency.

\subsection{Classification Performance}
As shown in Table~\ref{tab:test_metrics}, Q-DIVER achieved a Test F1-score of 63.49\%, exceeding the validation F1-score of 59.81\%, which suggests stable performance on the held-out test set.

\begin{table}[ht]
\centering
\caption{Classification performance metrics for the Q-DIVER model on the PhysioNet-MI dataset}
\label{tab:test_metrics}
\vspace{0.2cm}
\begin{tabular}{@{}lccccc@{}}
\toprule
\textbf{Split} & \textbf{Accuracy} & \textbf{F1-Score} & \textbf{Kappa} & \textbf{Precision} & \textbf{Recall} \\ \midrule
Validation & 59.77\% & 59.81\% & 0.4636 & 59.89\% & 59.77\% \\
\textbf{Test} & \textbf{63.39\%} & \textbf{63.49\%} & \textbf{0.5118} & \textbf{63.91\%} & \textbf{63.38\%} \\ \bottomrule
\end{tabular}
\end{table}



\subsection{Optimal Quantum Architecture Discovery}
Q-DIVER successfully navigated the search space of 576 potential architecture combinations (24 candidates per component) to identify a task-optimal circuit configuration. The search process converged on specific structural motifs that maximize expressivity for EEG signal classification.

\begin{figure}[!ht]
    \centering
    \begin{subfigure}[ht]{0.5\textwidth}
        \centering
        \includegraphics[height=1.9in]{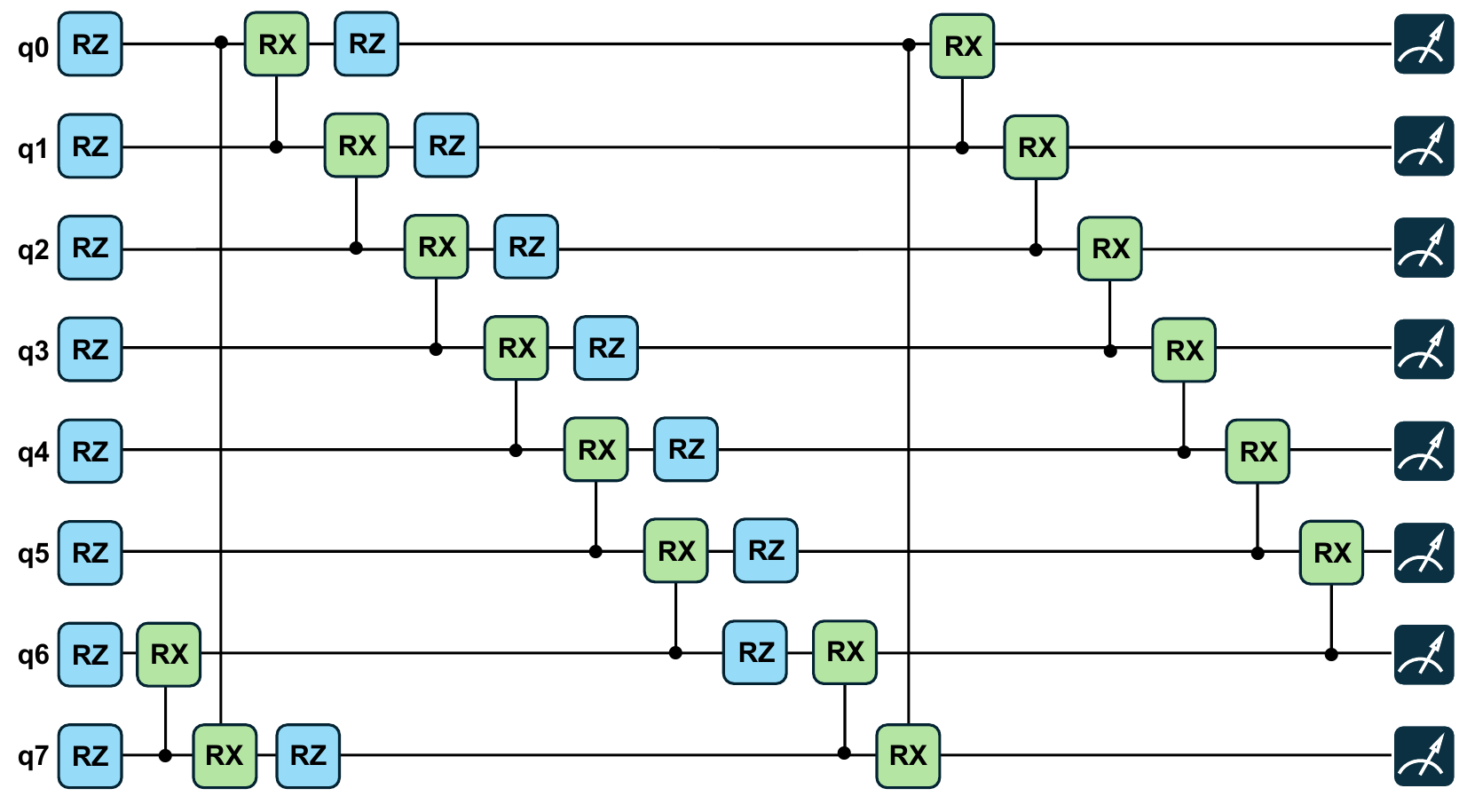}
        \caption{Timestep Circuit}
        \label{Figure_Timestep}
    \end{subfigure}%
    \hfill    
    \begin{subfigure}[ht]{0.5\textwidth}
        \centering
        \includegraphics[height=1.8in]{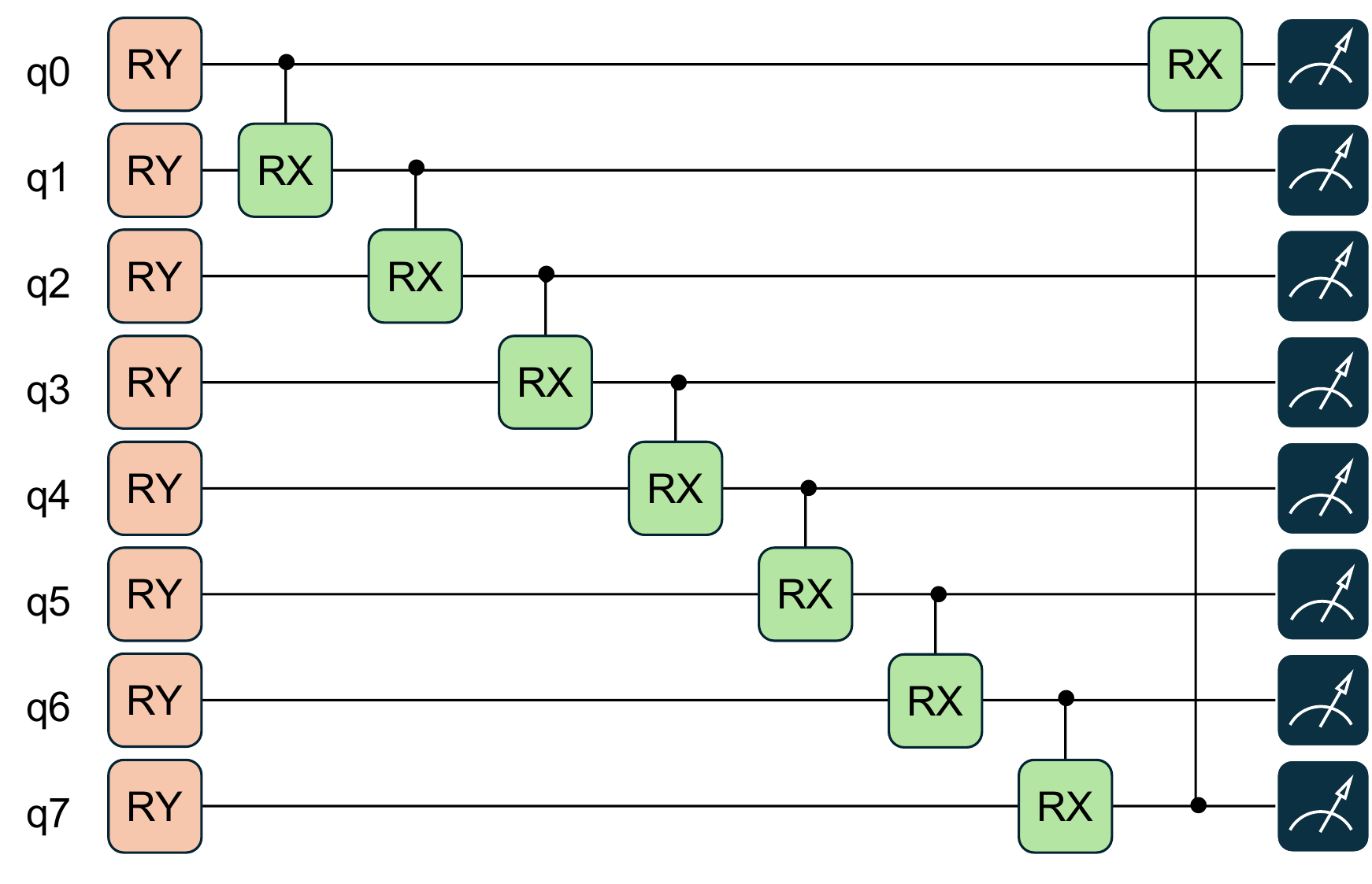}
        \caption{Quantum Feed Forward}
        \label{Figure_QFF}
    \end{subfigure}
    \caption{Optimal Quantum Circuit Architecture Obtained from DiffQAS.}
    \label{Fig_Ansatz}
\end{figure}

\begin{itemize}
    \item \textbf{Timestep Circuit:} The search selected a 2-layer architecture utilizing a \textbf{CRX Backward Ring} entangling pattern and \textbf{RZ} variational gates (Fig. \ref{Figure_Timestep}).
    \item \textbf{Quantum Feed-Forward (QFF) Circuit:} The search selected a 1-layer architecture utilizing a \textbf{CRX Forward Ring} entangling pattern and \textbf{RY} variational gates (Fig. \ref{Figure_QFF}).
\end{itemize}

Three distinct patterns emerged from the optimal architecture:
\begin{itemize}
    \item \textbf{Preference for Parametric Entanglement:} Both optimal circuits selected CRX gates over static CNOT gates. This suggests that the additional learnable parameters in the entangling layers are crucial for capturing the complex correlations in high-dimensional EEG data.
    \item \textbf{No Initialization Layer:} Neither the timestep nor the QFF circuit selected the Hadamard initialization layer. This indicates that for this specific motor imagery task, starting from the computational basis state $|0\rangle^{\otimes n}$ and relying solely on variational rotations provides sufficient expressivity.
    \item \textbf{Complementary Variational Gates:} The pipeline utilizes complementary rotation axes---RZ (phase) for temporal modeling and RY (amplitude) for feature extraction---enhancing the diversity of transformations within the quantum Hilbert space.
\end{itemize}

\subsection{Architecture Weight Convergence}
The differentiable search process exhibited stable convergence behavior. Initialized uniformly at $w = 1/24 \approx 0.0417$ during the 5-epoch warmup phase, the architecture weights for the optimal circuits gradually increased throughout training. By epoch 99, the weights for the optimal Timestep and QFF circuits reached approximately \textbf{0.063} and \textbf{0.065}, respectively. This gradual divergence, rather than a sharp winner-takes-all collapse, suggests that the soft selection mechanism allows the model to robustly explore the search space before settling on the optimal configuration.

\subsection{Parameter Efficiency}
For PhysioNet-MI with 4-second motor imagery trials (8 qubits), DIVER+QTSTransformer has 53.46M parameters (51.36M backbone + 2.10M head), whereas DIVER+MLP has 156.38M (51.36M + 105.02M). With end-to-end fine-tuning (unfrozen backbone), this yields an overall reduction of $\mathbf{2.9\times}$ at the full-model level.

Focusing on the task-specific readout head, QTSTransformer uses 2.10M parameters versus 105.02M for the MLP head, corresponding to an approximate $\mathbf{50\times}$ reduction. Most head parameters in the quantum configuration come from the classical feature projection (2.097M); the quantum circuit contributes 43 trainable parameters and the post-measurement linear classifier adds 100. This compact decision head suggests that the optimized quantum ansatz can provide an expressive readout under severe parameter constraints.

\section{Conclusion}
In this paper, we introduced Q-DIVER, a hybrid framework synergizing the DIVER-1 foundation model with the QTSTransformer to enable quantum transfer learning for high-dimensional EEG data. By extending DiffQAS, our approach autonomously discovered task-optimal circuit topologies, eliminating the reliance on heuristic ansatz design.

Empirical results on the PhysioNet-MI dataset highlight three key contributions. First, the quantum classifier matched classical MLP performance while using approximately $\mathbf{50\times}$ fewer task-specific head parameters, with detailed parameter comparisons provided in Table~\ref{tab:param_compare_breakdown} and the Results section.
This demonstrates that substantial parameter savings arise primarily from the decision-making module, supporting the high effective dimension of quantum feature spaces \cite{Abbas}. 
Second, the architecture search consistently converged on expressive parametric entangling gates (CRX) and eschewed initialization layers, revealing distinct, task-specific inductive biases \cite{DiFFQAS, DiffQAS_QLSTM}. 
Finally, the model exhibited robust generalization with test performance exceeding validation metrics (Test F1: 63.49\%), consistent with theoretical predictions regarding the favorable generalization bounds of constrained quantum models 
\cite{Caro2022, Gil-Fuster, Wu_Lorenz}.

Q-DIVER offers a viable pathway for deploying powerful, interpretability-friendly neuroimaging models in resource-constrained environments. Beyond empirical performance, our findings cautiously suggest that the structural constraints of quantum circuits may induce a task-relevant inductive bias, enabling effective decision boundaries under severe parameter limitations. Future work will expand this framework to diverse neurological domains and integrate quantum error mitigation to enhance robustness on NISQ hardware.

\section*{Acknowledgment}

Special thanks to the members of the SNU Connectome Lab, particularly Taeyang Lee, for their invaluable support to implementing hybrid quantum transfer learning of the DIVER-1 model. This work was supported by the National Research Foundation of Korea (NRF) grant funded by the Korea government (MSIT) (No. 2021R1C1C1006503, RS-2023-00266787, RS-2023-00265406, RS-2024-00421268, RS-2024-00342301, RS-2024-00435727, NRF-2021M3E5D2A01022515, and NRF-2021S1A3A2A02090597), by the Creative-Pioneering Researchers Program through Seoul National University (No. 200-20240057, 200-20240135). Additional support was provided by the Institute of Information \& Communications Technology Planning \& Evaluation (IITP) grant funded by the Korea government (MSIT) [No. RS-2021-II211343, 2021-0-01343, Artificial Intelligence Graduate School Program, Seoul National University] and by the Global Research Support Program in the Digital Field (RS-2024-00421268). This work was also supported by the Artificial Intelligence Industrial Convergence Cluster Development Project funded by the Ministry of Science and ICT and Gwangju Metropolitan City, by the Korea Brain Research Institute (KBRI) basic research program (25-BR-05-01), by the Korea Health Industry Development Institute (KHIDI) and the Ministry of Health and Welfare, Republic of Korea (HR22C1605), and by the Korea Basic Science Institute (National Research Facilities and Equipment Center) grant funded by the Ministry of Education (RS-2024-00435727). We acknowledge the National Supercomputing Center for providing supercomputing resources and technical support (KSC-2023-CRE-0568). An award for computer time was provided by the U.S. Department of Energy’s (DOE) ASCR Leadership Computing Challenge (ALCC). This research used resources of the National Energy Research Scientific Computing Center (NERSC), a DOE Office of Science User Facility, under ALCC award m4750-2024, and supporting resources at the Argonne and Oak Ridge Leadership Computing Facilities, U.S. DOE Office of Science user facilities at Argonne National Laboratory and Oak Ridge National Laboratory.

\clearpage

\bibliographystyle{ieeetr}
\bibliography{references}

\end{document}